\documentclass[a4paper,10pt,twoside]{cpc-hepnp}

\usepackage{fancyhdr}            %页眉和页码
\usepackage{multicol}            %用于双栏排版
\usepackage{upgreek}             %希腊字母的输入
\usepackage{indentfirst}         %支持首行缩进
\usepackage{booktabs}            %画三线表
\usepackage{array,tabularx}      %支持各种表格
\usepackage{keyval,graphicx}     %用于插图
\usepackage{textcomp}            %各种符号的输入
\usepackage{amssymb,bm,mathrsfs,bbm,amscd} %各种数学符号的输入
\usepackage[tbtags]{amsmath}     %AMSLaTeX 数学公式的录入，[tbtags]用于公式编号与公式的最后一行对齐
\usepackage{lastpage}            %计算总页数
\usepackage[numbers,sort&compress]{natbib}  %参考文献的上标
\usepackage{color}

\begin{document}

%===================================================================================================================
\setlength{\abovecaptionskip}{4pt plus1pt minus1pt}   %图形中的图与标题之间的距离
\setlength{\belowcaptionskip}{4pt plus1pt minus1pt}   %表格中的表与标题之间的距离
\setlength{\abovedisplayskip}{6pt plus1pt minus1pt}   %公式前的距离
\setlength{\belowdisplayskip}{6pt plus1pt minus1pt}   %公式后面的距离
\addtolength{\thinmuskip}{-1mu}            %减小自变量与函数关系的间距，比如sin x
\addtolength{\medmuskip}{-2mu}             %减小+,-,\times,\div的两侧空白
\addtolength{\thickmuskip}{-2mu}           %减小等号、不等号的两侧空白
\setlength{\belowrulesep}{0pt}          %在使用booktabs宏包画三线表中，加了 | 来加竖线时，会使横线和竖线不能很好
\setlength{\aboverulesep}{0pt}          %的交叉，使用这两个设置即可。
\setlength{\arraycolsep}{2pt}           %减小eqnarray中＝号两边的距离

\providecommand{\e}[1]{\ensuremath{\times 10^{#1}}}

%===================================================================================================================

\fancyhead[c]{\small Chinese Physics C~~~Vol. 37, No. 5 (2013) 057004}
\fancyfoot[C]{\small 057004-\thepage}

\footnotetext[0]{Received 10 July 2012}%, Revised 4 July 2012}

\title{\boldmath Thermal analysis and cooling structure design of the primary collimator in CSNS/RCS}

\author{%
ZOU Yi-Qing(邹易清)$^{1,2;1)}$\email{zouyq\oa ihep.ac.cn}
\quad WANG Na(王娜)$^{1}$
\quad KANG Ling(康玲)$^{1}$\\
QU Hua-Min(屈化民)$^{ 1}$
\quad HE Zhe-Xi(何哲玺)$^{1,2}$
\quad YU Jie-Bing(余洁冰)$^{1}$%
}

\maketitle

\address{%
$^{1}$ Institute of High Energy Physics, Chinese Academy of Sciences, Beijing 100049, China\\
$^{2}$ Graduate University of Chinese Academy of Sciences, Beijing 100049, China\\
}

\begin{abstract}
The rapid cycling synchrotron (RCS) of {the} China Spallation
Neutron Source (CSNS) is a high intensity proton ring with beam power of
100~kW. In order to control the residual activation to meet the
requirements of hands-on maintenance, a two-stage collimation system has
been designed for the RCS. The collimation system consists of
one primary collimator made of thin metal to scatter the beam and four
secondary collimators as absorbers. Thermal analysis is an important aspect
in evaluating the reliability of the collimation system. The calculation of
the temperature distribution and thermal stress of the primary collimator
with different materials is carried out by using ANSYS code. In order to
control the temperature rise and thermal stress of the primary collimator to
a reasonable level, {an} air cooling structure is intended to be used. The
mechanical design of the cooling structure is presented, and the cooling
efficiency with different chin numbers and wind velocity {is} also analyzed.
Finally, the fatigue lifetime of the collimator under thermal shocks is estimated.
\end{abstract}


\begin{pacs}
29.27.Eg, 44.05.+e, 44.10.+i \qquad {\bf DOI:} 10.1088/1674-1137/37/5/057004
\end{pacs}

\footnotetext[0]{\hspace*{-3mm}\raisebox{0.3ex}{$\scriptstyle\copyright$}2013
Chinese Physical Society and the Institute of High Energy Physics
of the Chinese Academy of Sciences and the Institute
of Modern Physics of the Chinese Academy of Sciences and IOP Publishing Ltd}%

\begin{multicols}{2}

\section{Introduction}

In high intensity proton rings, beam induced activation is a primary concern
during the design of the accelerators. To achieve hands-on maintenance, the
beam loss should be well controlled below 1~W/m, which is hard to achieve
during the design and construction of an actual machine. {A} beam collimation
system is often used to concentrate the beam loss in a restricted area to
reduce beam loss deposition in other areas around the ring [1].

The CSNS will provide beam pulses of 1.56$\times $10$^{13}$ particles with
repetition rate of 25~Hz. The RCS is designed to accelerate the proton beam
from 80~MeV to 1.6~GeV [2]. A two-stage collimation system will be used
for the transverse beam collimation in the RCS. Halo particles with large
amplitude will first hit the primary collimator. After interaction, the
particles will be scattered by the primary collimator without being
absorbed. Then, after some certain distance, the particles will finally be
absorbed by the secondary collimators downstream {from} the primary one [3, 4].

The primary collimator consists of four scrapers, which are set either
horizontally or vertically as shown in Fig.~1. Each scraper has a dimension
of 150~mm$\times $30~mm ($W\times H)$ in the transverse plane. Movable
collimators with air cooling are chosen for the transverse beam collimation
in CSNS. This piece is brazed to an air-cooled copper block. Fig.~4 shows
{the} detail of the scraper mechanism. The scraper slice, bellow, slide and
cooling fin can be seen.

%fig 1
\begin{center}
\includegraphics{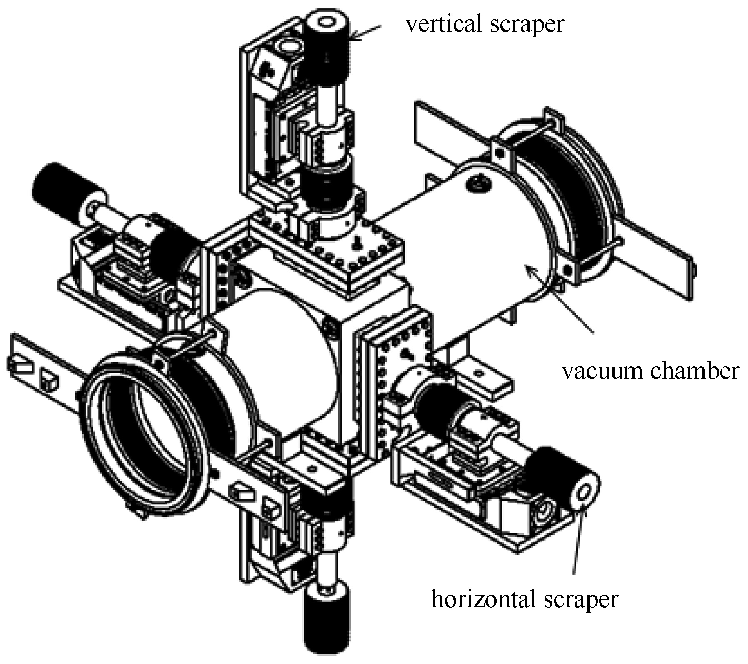}
\figcaption{Schematic view of the primary collimator.}
\end{center}

Three kinds of metal, copper, tungsten and tantalum, have been considered as
material of the primary collimator for their high melting point and good
thermal conductivity. The thicknesses of the materials are optimized for
providing efficient scattering angles and high collimation efficiency. The
parameters of different materials are listed in Table~1.

%table 1
\begin{center}
\renewcommand{\arraystretch}{1.2}
\tabcaption{Three materials scraper energy and power loss.} \footnotesize
\begin{tabular*}{86mm}{@{\extracolsep{-1mm}}cccc}
\toprule
materials & energy loss & power & Max power\\
thickness/mm & $\Delta E$/MeV & deposition/N(W) & density/(W/m$^{2})$ \\
\hline
Ta(0.2~mm)&
1.23&
102.3&
7.915e6 \\
W(0.17~mm)&
1.24&
103.2&
7.985e6 \\
Cu(1.0~mm)&
4.38&
364.4&
28.21e6 \\
\bottomrule
\end{tabular*}%
\end{center}

In order to provide {an} adequate scattering angle, the scrapers need to be very
thin ($<$1~mm), which results in high energy deposition. Excessive
temperature rise or inhomogeneous temperature distribution induced thermal
stress effects may cause serious damage to the collimator. So the thermal
analysis and properly designed cooling structure are necessary for the
reliability of the collimator.

In this article, we first present the energy deposition caused by beam loss
on the primary collimator. The expected time structure of the beam is taken
into account. Then thermal and mechanical analyses for different materials
are performed by using ANSYS code. Steady state heat transfer analysis is
performed to determine the temperature distribution. Based on the analysis,
the cooling structure is designed and optimized with chin numbers and wind
velocity. Finally, the fatigue lifetime under thermal shock is estimated
with transient thermal analysis.

\section{Energy deposition calculation}

The average power deposition on the primary collimator can be estimated by
the following formula
\begin{equation}
%1
\label{eq1}
\bar {P}({\rm W})=nef\Delta E,
\end{equation}
where $n$ is the number of particles {hitting} the primary collimator, $e$ is the
charge of an electron, $f$ is the repetition rate of the RCS, and $\Delta E$ is the
ionization energy loss of the particle during interaction with the collimator.

Assuming 10{\%} of the total beam will interact with the primary collimator.
The ionization energy loss and the average beam power deposition of
different materials in the collimator are shown in Table~1.

The beam loss distribution on the primary collimator is estimated by using
ORBIT code [5]. A realistic beam distribution obtained by painting
injection is tracked and accelerated from 80~MeV to 1.6~GeV in the presence
of space charge. We project the integrated beam losses in the {scraper} onto
the transverse plane, and fit the beam loss distribution with Gaussian
functions as shown in Fig.~2. The particle loss is mainly concentrated in the
range: $-$25~mm$<W<$25~mm, 0$<H<$3~mm. The corresponding average power
density distribution is expressed as
\begin{eqnarray}
%2
\label{eq2}
 P({\rm W}/{\rm mm}^2)&=&\frac{N}{2p\times 8.2}\exp \left( {-\frac{(x+0.7)^2}{2\times
8.2^2}} \right)\nonumber\\[4mm]
&&\times\left( \frac{28.4}{1.8}\exp \left( {-\frac{(y+3.3)^2}{2\times 1.8^2}}
\right)\right.\nonumber \\[4mm]
&&\left. +\frac{741.2}{0.5}\exp \left( {-\frac{(y+1.9)^2}{2\times 0.5^2}}
\right) \right),
\end{eqnarray}
where $N$ is the average power deposition in the first three milliseconds, $x$ and
$y$ are the transverse positions in the $W$ and $H$ direction in the transverse plane
with unit of mm, and the origin $x=y$=0 locates at the center of the edge close
to the beam.

%fig 2
\begin{center}
\includegraphics{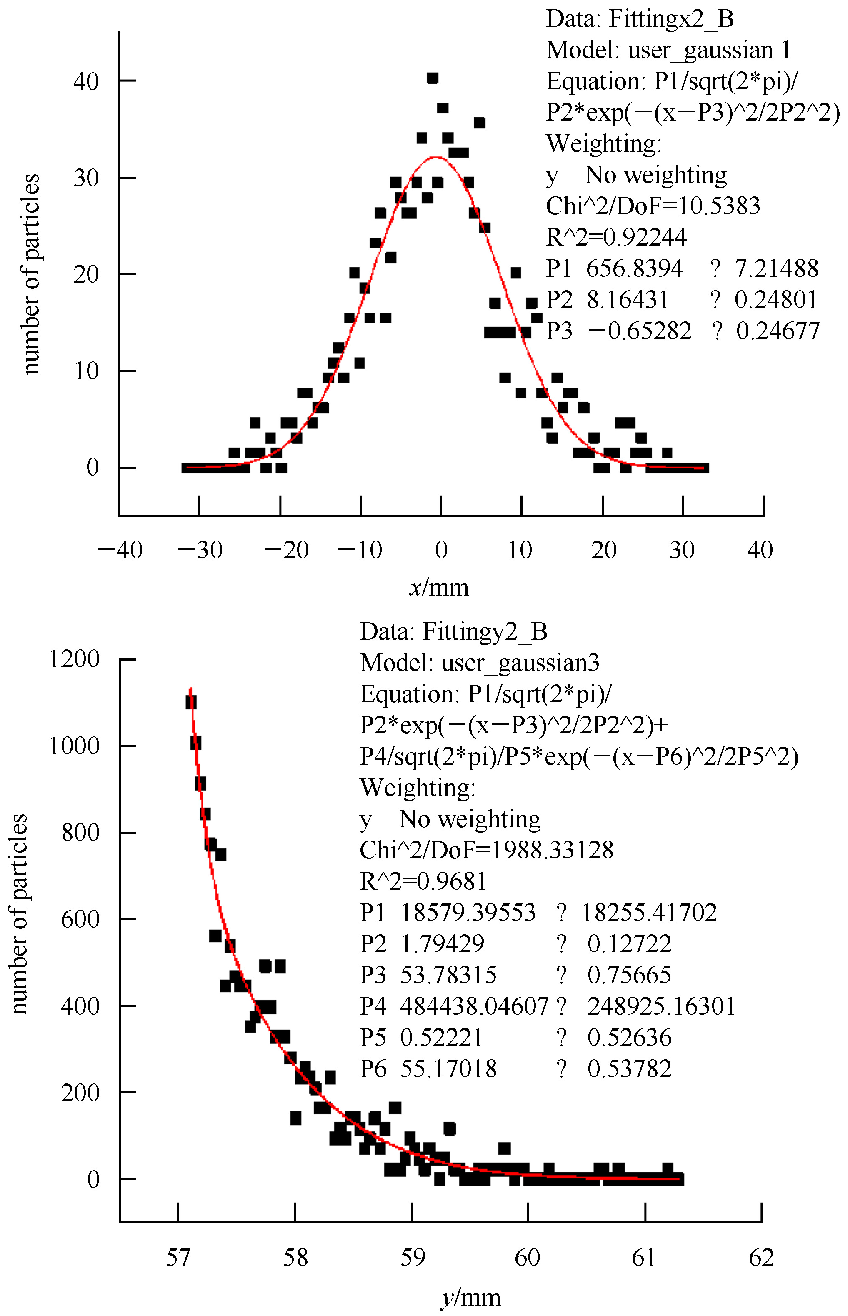}
\figcaption{Fitting of the beam loss distribution (Top: width, Bottom: height).}
\end{center}

The time structure of the beam loss distribution is also considered. Fig.~3
shows the beam power loss distribution versus time. According to the
simulations, the beam losses {mostly} occur during the first three milliseconds
with average beam energy of 100~MeV.

%fig 3
\begin{center}
\includegraphics{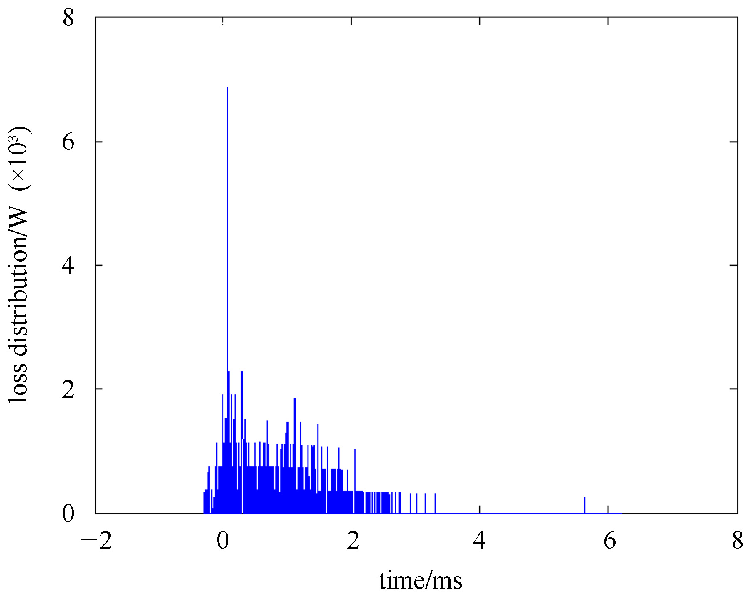}
\figcaption{The beam power density loss distribution vs. time.}
\end{center}

\section{Cooling structure design}

Figure~1 and Fig.~4 show the scheme of the scraper structure. Four primary
collimator scrapers are driven by four stepping motors individually. Each
scraper connects with a copper block and cooling fin by silver brazing.
{The} primary collimator will be set in the CSNS/RCS tunnel, and the cooling fin
is at room temperature. The cooling structure consists of 15 pieces of chins
with separation of 9~mm.

%fig 4
\begin{center}
\includegraphics{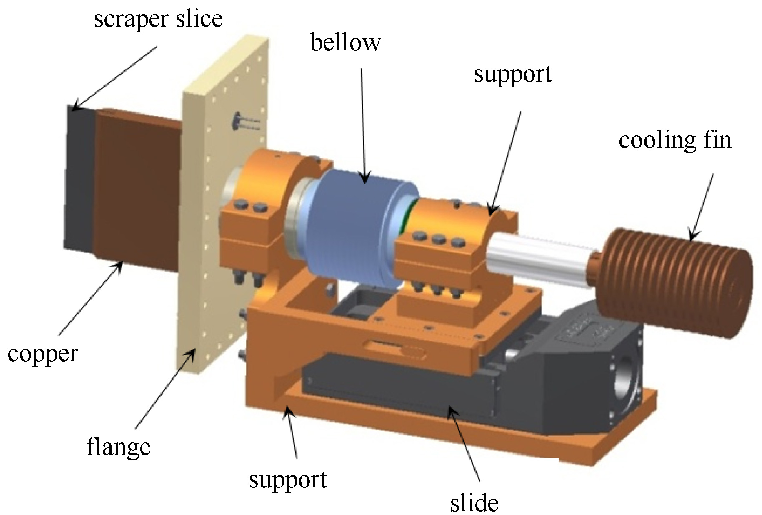}
\figcaption{The scheme of the scraper structure.}
\end{center}

\section{Steady-state thermal analysis}

Thermal and mechanical FEM analyses are carried out with ANSYS for
investigating the heat distribution and thermal stress and deformation.

\subsection{Heat transfer analysis}

According to the characteristic of the energy deposition, steady-state heat
transfer analysis is performed to determine the temperature distribution.
Convective heat transfer boundary condition is used. The FEM model of one
scraper is shown in Fig.~5. Free air cooling condition has been considered.
The maximum temperatures of different materials are presented in Table~2.
From the result we can see that copper gives the lowest temperature rise,
and tantalum presents the highest, but they are all well below the melting points.

%fig 5
\begin{center}
\includegraphics{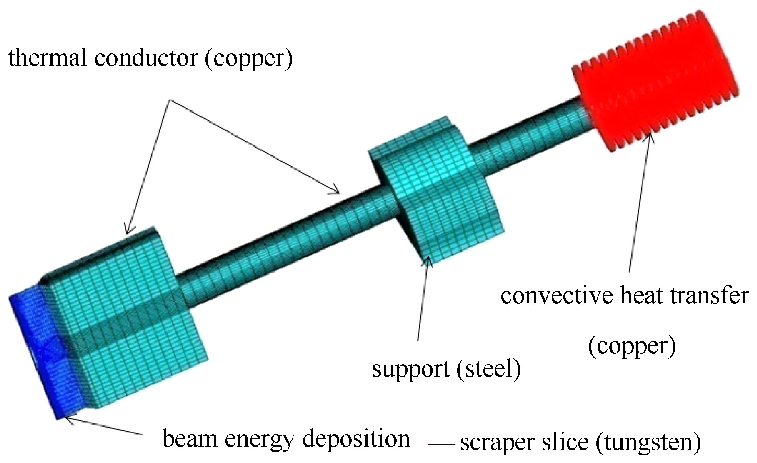}
\figcaption{The scraper FEM model and loads.}
\end{center}

%table 2
\begin{center}
\renewcommand{\arraystretch}{1.2}
\tabcaption{The analysis results of different materials.} \footnotesize
\begin{tabular*}{86mm}{@{\extracolsep{\fill}}cccc}
\toprule
material & Max & Max Von mises & Max\\
thickness & temperature/K & stress/MPa & strain/mm \\
\hline
Ta(0.2 mm)&
805.609&
187&
0.183 \\
W(0.17 mm)&
506.517&
168&
0.149 \\
Cu(1.0 mm)&
431.313&
315&
0.498 \\
\bottomrule
\end{tabular*}%
\end{center}

\subsection{Static stress calculation}

The temperature distributions are input in structural analyses to determine
the thermal stresses and displacements caused by the temperature loads. The
comparison among different materials is given in Table~2. We can see that
although copper presents the lowest temperature rise, the stress induced by
the heat load exceeds the allowable stress of copper and the deformation is
too large. Compared with tantalum, the temperature of the tungsten scraper
is relatively low; the maximum stress of tungsten scraper is lower and it
can get better mechanical properties; the deformation of tungsten is smaller
and it can get better precision. Based on the above analysis, tungsten is
chosen for the material of the CSNS/RCS primary collimator. {The}
analysis and simulation of the following chapters are for tungsten scrapers.
The temperature distribution of {the} tungsten scraper is shown in Fig.~6.
and Fig.~7 presents the Von Mises stress distribution of {the} tungsten scraper.

%fig 6
\begin{center}
\includegraphics{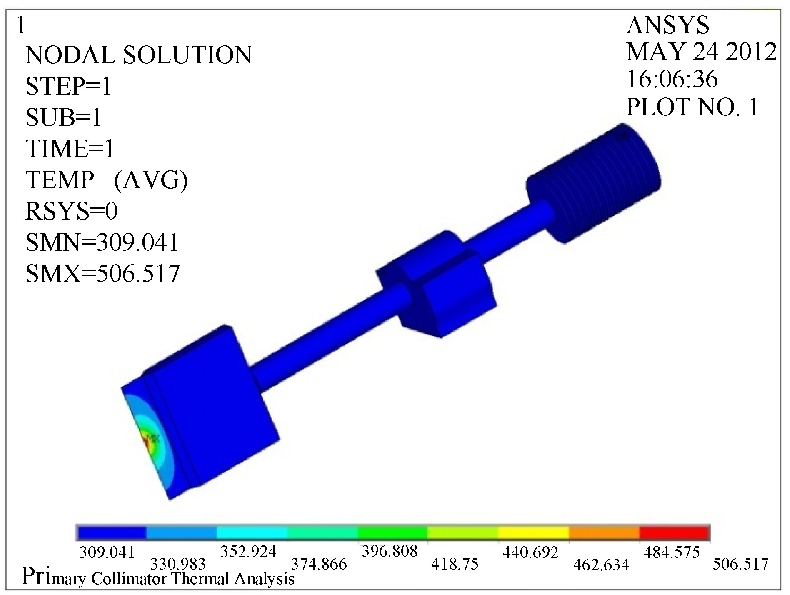}
\figcaption{The temperature distributions of the tungsten scraper.}
\end{center}

%fig 7
\begin{center}
\includegraphics{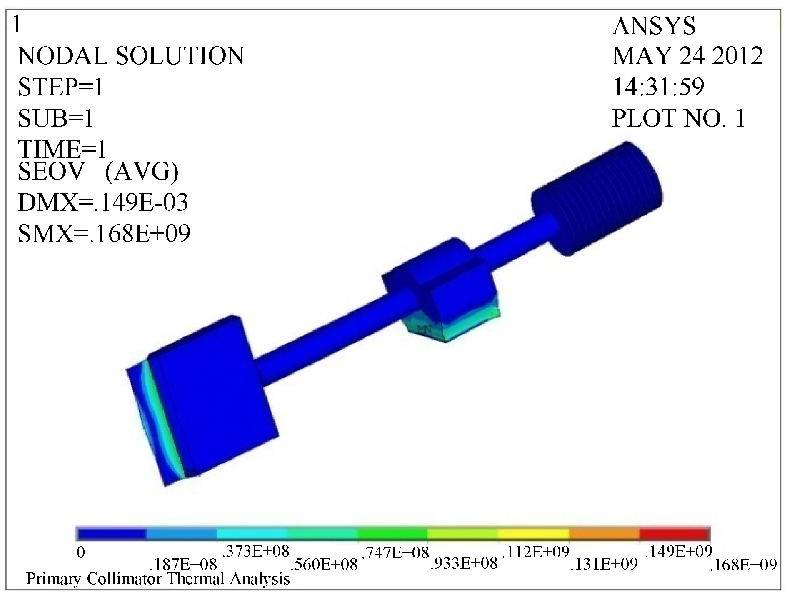}
\figcaption{The Von Mises stress distributions of the tungsten scraper.}
\end{center}

The regions of {the interface of} different materials, the centralized heat source
and the fixed constraint support are the dangerous areas. These should be
carefully designed and taken {into serious} consideration in the
process of manufacture.

\section{Cooling efficiency}
\subsection{Cooling efficiency of different chin numbers}

There are a number of studies on the air-cooling of air-cooled engine fins.
The average fin surface heat transfer coefficient can be obtained using the
following equation [6]
\begin{equation}
%3
\label{eq3}
\alpha _{\rm avg} =\left(2.47-\frac{2.55}{p^{0.4}}\right)\mu ^{0.9}+0.0872p+4.31 ,
\end{equation}
where $p$ is the fin pitch, and $\mu $ is the wind velocity. The equation is valid
when $\mu $ is in the range of 0 to 60~km/h.

Assuming free air-cooling condition, we can get the maximum temperature of
scrapers with different fin numbers. The results are shown in Fig.~8.

%fig 8
\begin{center}
\includegraphics{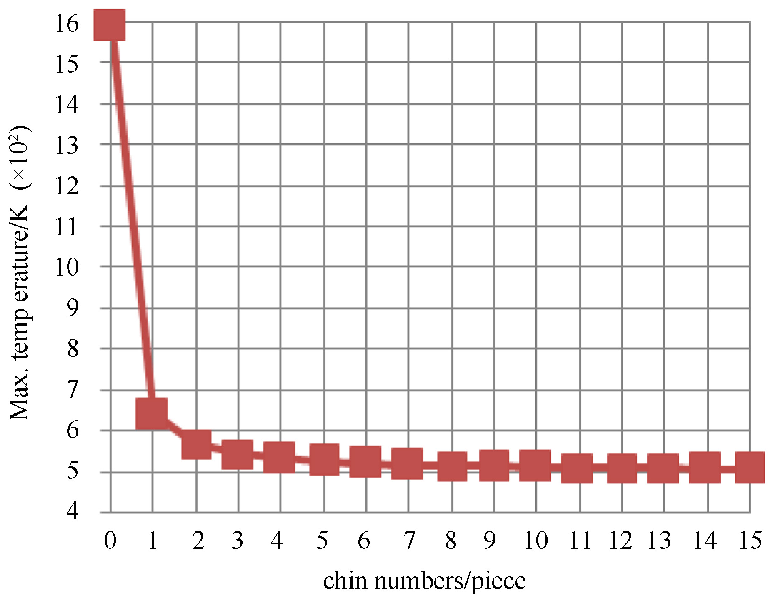}
\figcaption{Max temperature vs. chin numbers.}
\end{center}

As shown in the plot, the temperature drops dramatically due to the
introduction of cooling fins. With the increasing of the chin numbers, the
temperature saturates when the number of chin is larger than 5. The results
further prove that {an} air cooling structure is necessary for controlling the
temperature rise induced by {the} heat load. We can design or purchase optimum
air-cooled {fins} according to the results.

\subsection{Cooling efficiency of different wind velocity}

Generally, forced air cooling can greatly improve the air cooling effect.
But this is not the case in the thermal analysis of the primary collimator
in the RCS. The\linebreak\vspace{-1mm}

%fig 9
\begin{center}
\includegraphics{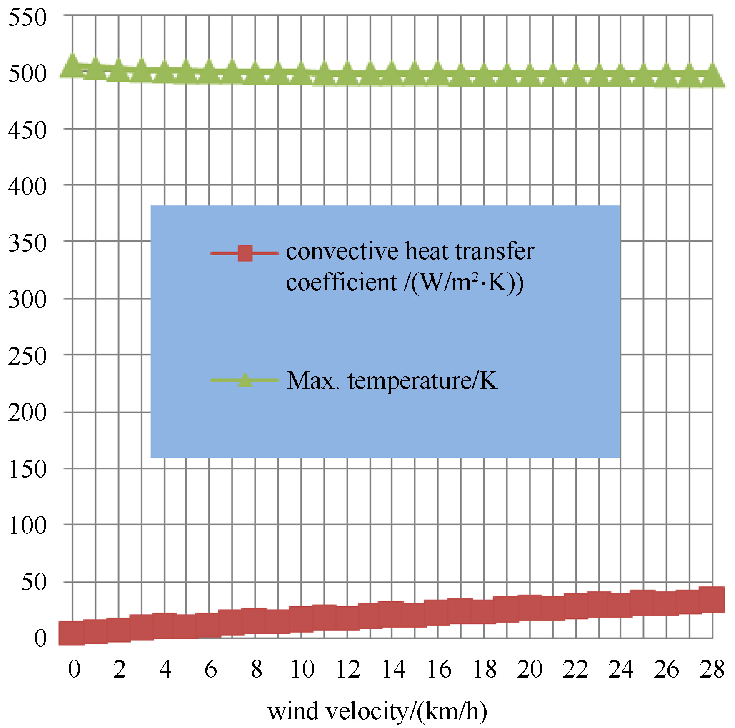}
\figcaption{Max temperature, heat transfer coefficient vs. wind velocity.\vspace{-2mm}}
\end{center}

\noindent temperature changed very {little} by increasing wind velocity
in our estimation. The relation of the maximum temperature and convective
heat transfer coefficient with the wind velocity is shown in Fig.~9. So the
forced air cooling system is not necessary in the CSNS/RCS primary
collimator cooling structure.

\section{Transient thermal analysis}

In order to simplify the calculation, an equivalent beam power deposition in
the first three milliseconds is assumed in the transient thermal analysis.
The temperature approximately reaches a steady state around 60000 s (about
1.5E6 cycles). The result is shown in Fig.~10. The temperature vibrates
between 487~K and 521~K in one period, which corresponds to a maximum
temperature difference of 34~K.

%fig 10
\vspace{2mm}
\begin{center}
\includegraphics{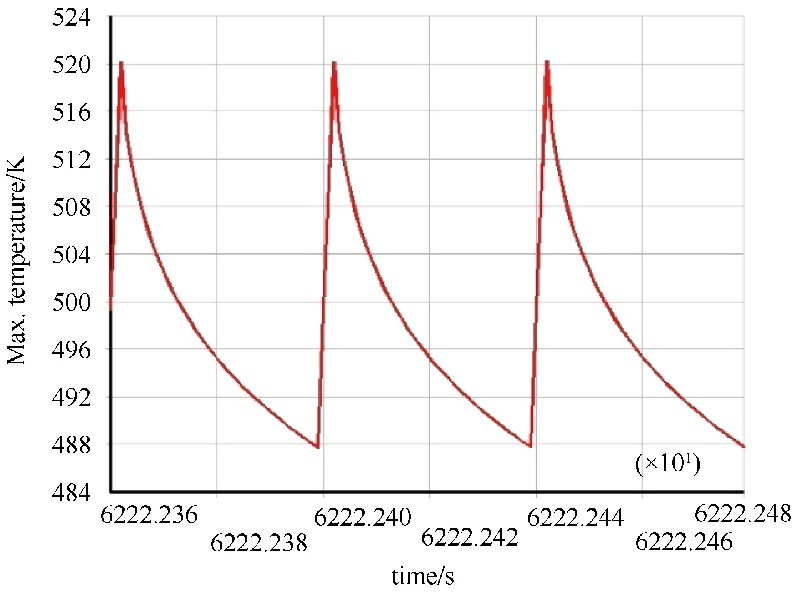}
\figcaption{The scraper static-state temperature versus time.}
\end{center}

Based on the transient thermal analysis, the thermal shock stress is
estimated. Fig.~11 shows the results of maximum period stress vs. time in the
first 10 periods.

The maximum Von Mises stress temperature difference within one period is
about 37~MPa. The fatigue limit stress can be expressed by
\begin{equation}
\sigma _0 =1.6Hv\pm 0.1Hv~{[7]},
\end{equation}
where $Hv$ is the Vickers hardness in kgf/mm$^{-2}$.

The Vickers hardness of tungsten in 506~K is about 100~kgf/mm$^{-2}$
[8]. The fatigue limit stress is about 160~MPa which is much larger than
the maximum period stress. So the scrapers are not supposed to suffer a
fatigue failure by thermal shock.

%fig 11
\begin{center}
\includegraphics{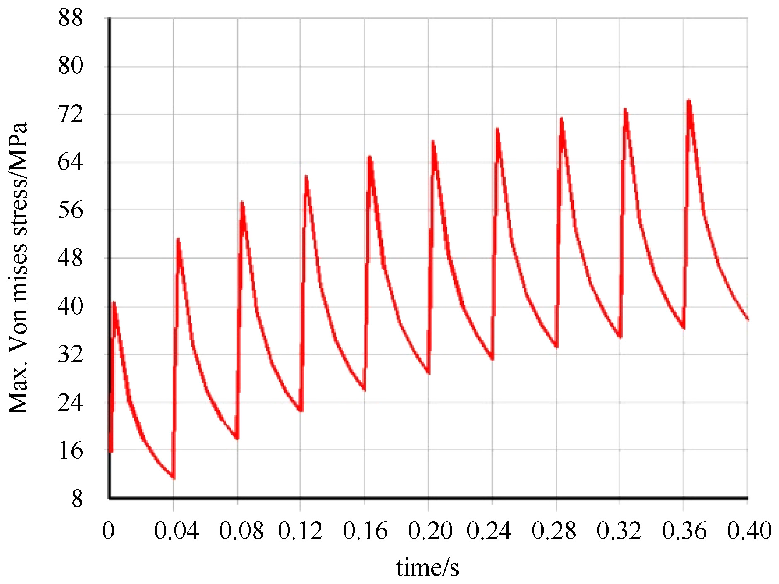}
\figcaption{The scraper maximum Von Mises stress versus time.}
\end{center}

\section{Conclusions}

Through the above analysis and simulation, we can draw the following
conclusions. (\ref{eq1}) From the simulation of the thermo-mechanical properties of
three different materials, tungsten is finally chosen as the material of the
CSNS/RCS primary collimator for its lower temperature rise and better
mechanical performance. (\ref{eq2}) The air-cooled fin is necessary and the forced
air cooling system is not needed in the CSNS/RCS primary collimator cooling
structure. (\ref{eq3}) The scraper will not suffer a fatigue failure by thermal
shock from the transient thermo-mechanical analysis.

\end{multicols}

\vspace{-2mm}
\centerline{\rule{80mm}{0.1pt}}
\vspace{2mm}

\begin{multicols}{2}

%参考文献

\end{multicols}

\clearpage

\end{document}